# High-Efficient *ab initio* Bayesian Active Learning Method and Applications in Prediction of Two-dimensional Functional Materials

Xing-Yu Ma,[1†]  Hou-Yi Lyu,[1,2†]  Kuan-Rong Hao,[1]  Zhen-Gang Zhu,[3,1]  Qing-Bo Yan,[2*] Gang Su[4,1,2*]

[1]School of Physical Sciences, University of Chinese Academy of Sciences, Beijing 100049, China.
[2]Center of Materials Science and Optoelectronics Engineering, College of Materials Science and Optoelectronic Technology, University of Chinese Academy of Sciences, Beijing 100049, China.
[3]School of Electronic, Electrical and Communication Engineering, University of Chinese Academy of Sciences, Beijing 100049, China.
[4]Kavli Institute for Theoretical Sciences, and CAS Center of Excellence in Topological Quantum Computation, University of Chinese Academy of Sciences, Beijing 100190, China.

**Beyond the conventional trial-and-error method, machine learning offers a great opportunity to accelerate the discovery of functional materials, but still often suffers from difficulties such as limited materials data and unbalanced distribution of target property. Here, we propose the *ab initio* Bayesian active learning method that combines active learning and high-throughput *ab initio* calculations to accelerate prediction of desired functional materials with the ultrahigh efficiency and accuracy. We apply it as an instance to a large family (3,119) of two-dimensional hexagonal binary compounds with unbalanced materials property, and accurately screen out the materials with maximal electric polarization and proper photovoltaic band gaps, respectively, whereas the computational costs are significantly reduced by only calculating a few tenths of possible candidates in comparison to the random search. This approach shows enormous advantages for the cases with unbalanced distributions of target property. It can be readily applied to seek for a broad range of advanced materials.**

# Introduction

Designing functional materials with desired properties is an important frontier in materials science. With traditional trial-and-error methods, a huge high-dimensional search space of materials with tremendous structural and chemical complexities and enormous combinatorial possibilities makes it extremely difficult to discover the desired functional materials. The high-throughput *ab initio* calculations based on density functional theory (DFT)[1] were adopted to search for promising materials,[2,3,4] which requires to perform abundant DFT calculations in searching for few desired materials from thousands or even more candidates at a low efficiency. To overcome these challenges, machine learning methodologies have been widely applied in materials and chemical sciences, [5-7] such as for predicting properties of inorganic crystals [8] and chemical reactivity, [9] designing lithium batteries[10] and catalysts, [11] and predicting two-dimensional (2D) optoelectronic materials[12] and 2D ferromagnetic materials, [13] *etc*. These advances demonstrate that supervised machine learning techniques could accelerate the prediction of new functional materials. However, the accuracy of supervised machine learning model is often hindered by the limited materials data and unbalanced distributions of target property, i.e., uneven instances for different labeled category in the materials data set. More importantly, for the imbalance problem, it is usually difficult to improve the accuracy of learning model by simply enlarging the size of data. Recently, it is shown that the active learning techniques may alleviate this issue and have several applications in designing ultranarrow-band thermal radiator, [14] predicting homobenzylic ether molecules, [15] layered materials, [16] and piezoelectric compounds,[17] synthesizing ternary alloys, [18] and generating machine-learning force fields, [19] *etc*. This approach could find an optimal value relying on the use of predictions from a model together with an acquisition function that iteratively prioritizes the decision-making process on next unexplored data.

Here, we propose an efficient method named as *ab initio* Bayesian active learning that combines the active learning and high-throughput *ab initio* calculations to effectively and accurately accelerate the prediction of desired functional materials from a complex high-dimensional search space comprised of thousands of materials.

As an example, it is applied to accurately predict several materials with maximal electric polarization and proper photovoltaic band gaps, respectively, from a large family of two-dimensional (2D) hexagonal binary compounds. It is found that the present computational costs are several ten times less than those of the random search. Meanwhile, this method also shows outstanding ability to predict the desired functional materials in the case of unbalanced distributions of target property. Besides, several stable novel ferroelectric and photovoltaic semiconductors with out-of-plane electric polarization are screened out. Our strategy can be generally applicable to predict various classes of desired functional materials using a minimal number of calculated candidates, especially for unbalanced distributions of target property.

## Results and discussion

### *Ab initio* Bayesian active learning method

Active learning is a machine learning method based on an adaptive scheme, which is used to iteratively guide selection of the next unexplored data using the acquisition function. [20] Then, these next new data are added in training data for retraining new model until we are satisfied with the current outcome. The critical key of the strategy is the acquisition function based on the predictions from a surrogate machine learning model. Here, we utilized active learning method to predict functional materials, specifically, materials data are generated by the high-throughput *ab initio* calculations and we adopt the Gaussian process regression (GPR) [21] as a surrogate model and its acquisition function is based on the uncertainties and the mean predicted values obtained by the GPR model. Thus, we presented a new method called the *ab initio* Bayesian active learning for accelerated prediction of functional materials. There are many available acquisition functions, such as the probability of improvement (PI), the upper confidence bounds (UCB), and the expected improvement (EI), *etc*. However, UCB has adjustable hyper parameters and PI tends to optimize data locally. The method was frequently used to optimize the high-dimensional black box functions with many variables by employing minimal function evaluations. [22] EI was widely adopted because it does not have adjustable hyper parameters and tends to optimize data globally compared with the two other acquisition functions. [20, 22] It has the following form to identify the optimal values

$$EI(x) = \sigma(x)[z\Phi(z) + \varphi(z)] \qquad (1)$$

where $z = (\mu(x) - f^{max})/\sigma(x)$ and $z = (f^{min} - \mu(x))/\sigma(x)$ for optimizing the maximum and minimum values, respectively, $f^{max}$ and $f^{min}$ are the maximum and minimum actual values observed so far in the training data set, respectively, $\mu(x)$ and $\sigma(x)$ are the average predicted output values and standard deviation values (uncertainties) of $\mu(x)$ for the input feature x in unexplored data set, respectively, which are all obtained by GPR model, $\Phi(z)$ and $\varphi(z)$ are the cumulative density function and probability density function of standard normal distributions, respectively. The acquisition function is comprised of two parts. The first part ($\sigma(x)z\Phi(z)$) finds a better value than the current best estimates (the maximum or minimum properties in the training data set), denoting an exploitation process. The second part ($\sigma(x)\varphi(z)$) is to explore the total search space where the uncertainties are the largest, representing an exploration process. Although the exploration and exploitation could not be maximized simultaneously, the maximum EI is a reasonable trade-off between them. Thus, the next new data with maximum EI are selected from the unexplored data set, and then added into the training data set for retraining GPR model. This adaptive process is iteratively executed until optimal results are obtained. Noting that the number of iterations and selecting initial data set could affect the discovery of the optimal values. If the results of multiple independent optimization runs are almost the same, it implies that the global optimal value is achieved. Contrariwise, they may be the local optimal values, and in this latter case, it is necessary to increase the number of iterations until the results of multiple runs are consistent. The advantage of the method is that it could evaluate few sampling data to identify the optimal value in a complex high-dimensional search space. In principle, the above method depends only on the acquisition function and does not involve specific materials and properties. Moreover, the EI maximizing process could select new candidates with better property as much as possible at each iteration. The Bayesian active learning method could enlarge the data size of less category and make the distributions of target property more balanced in the training data set. Thus, it is irrelevant to the distributions of materials data and could be applied to the prediction of desired materials in the cases of unbalanced distributions of target property. For discovery of a broad range of advanced materials, we only need to select the corresponding material features, such as classical force-field inspired features for thermoelectric materials,[23] modified X-ray diffraction features for lithium battery[24] and catalyst descriptors for catalysts.[25] Therefore, the *ab initio* Bayesian active learning could be generally applicable to the projection of a broad of advanced functional materials with desired properties in thousands of possible candidates, particularly in the cases with unbalanced distribution of target property.

The proposed scheme of the *ab initio* Bayesian active learning method is presented in Fig. 1. First, a small size of the total unlabeled data set is chosen as the initial data set. Second, the target property of materials in the initial data set is obtained by DFT calculations. The DFT-calculated property of materials in the initial data set usually falls short of the desired values. Third, the initial data set is all adopted as the training data set, which is comprised of materials features and the target property. We train a machine learning model by the GPR algorithm, which establishes a mapping relationship between the features and property of materials. The target property can be described by using the Gaussian process, which has the form of $GP(\mu(x), \sigma(x))$ where Gaussian process (GP) is a collection of random variables satisfying a joint Gaussian distribution, x is the input feature vectors of materials data, $\mu(x)$ and $\sigma(x)$ are their mean predicted properties and standard deviation values (uncertainties) of the joint Gaussian distribution, respectively. Fourth, the GPR model is used to obtain the average predicted target property $\mu(x)$ and the uncertainty $\sigma(x)$ of each material in the remaining data set. Then the EI acquisition function value for each of the remaining data set is calculated by Eq. (1) based on their $\mu(x)$ and $\sigma(x)$, and then the candidate corresponding to the maximal EI is selected. The candidate (the red star in Fig. 1) possesses the largest EI (The height of the blue shaded area in Fig. 1 represents the uncertainty). Then the property of this new candidate is calculated by DFT, which is added to the training data set for retraining. A new GPR model is then trained with the updated training data set. This feedback iterative process is repeated until the materials with the desired DFT-calculated property values are predicted, which are taken as the promising candidates for further investigation. It is noted that the direction of searching for the desired materials is determined by the EI based on the predictions of GPR model, and the property of each predicted material is obtained by the DFT. In principle, the method only evaluates few sampling materials data to predict the desired materials in a huge search space, and its accuracy is completely consistent with that of DFT. DFT has been widely applied in the materials simulations, and the obtained materials properties usually agree well with the experimental measured values.[26-28] In addition, the above process is independent of materials prototype, specific properties and data distributions. Therefore, the *ab initio* Bayesian active learning presented here is generally suitable for the prediction of a wide range of desired advanced functional materials with low computational costs and high precision. In the following, we will apply this method as an instance to predict 2D ferroelectric materials with ultrahigh polarization and the photovoltaic materials with proper band gaps, and evaluate its efficiency and accuracy.

**Application in the prediction of 2D ferroelectric materials**

2D ferroelectric materials, including $In_2Se_3$ and other $III_2$-$VI_3$ compounds,[29] Group IV monochalcogenides,[30, 31] $M_IM_{II}P_2X_6$ ($M_I$, $M_{II}$ = metal elements, X = O/S/Se/Te),[32-34] have been intensively studied due to a wide range of promising applications in electromechanical transducer,[35] ferroelectric field effect transistor[36] and ferroelectric tunnel junctions,[37] *etc*. Recently, 2D hexagonal binary compounds (HBCs) with hexagonal honeycomb structures attracted extensive attention, as monolayer BN, $GdAg_2$, $GdAu_2$ and $MgB_2$ have been successfully synthesized experimentally,[38-40] and monolayer NbN,[41] CrN and $CrB_2$[42] have also been predicted to be potential 2D ferroelectrics. Here we focus on the whole family of 2D HBCs and try to seek for the ferroelectric members with maximal polarization by employing the *ab initio* Bayesian active learning method. In addition, the computational costs and prediction accuracy are also examined.

There are two types of structures of 2D HBCs (MX and $MX_2$, M and X are different atoms): (i) MX shown in Fig. 2(a), a honeycomb lattice is formed by staggeringly arranging M atoms (yellow balls) and X atoms (purple balls), and (ii) $MX_2$ shown in Fig. 2(e), X atoms are arranged in a honeycomb lattice, in which M atoms are embedded. When M and X atoms locate on the same plane, the structures MX and $MX_2$ have space groups $P\bar{6}m2$ (No. 187) and *P6/mmm* (No.191) (or *Cmmm* (No.65) for some $MX_2$), respectively, corresponding to the non-polar point groups $\bar{6}m2$ and *6/mmm* (or *mmm*), respectively (Figs. 2(c) and (g)). They belong to the paraelectric phase as confirmed by our calculations. Meanwhile, when M and X atoms deviate from the original honeycomb plane with *d* the vertical distance between M and X, the space groups of MX and MX$_2$ reduce to *P3m1* (No. 156) and *P6mm* (No. 183) (or *Cmm2* (No. 35)), respectively, corresponding to the polar point groups *3m* (Figs. 2(b) and (d)) and *6mm* (or *mm2*) (Figs. 2(f) and (h)), respectively, and they belong to the ferroelectric phase. A number of 2D MX and MX$_2$ structures are generated in our practice by replacing the M and X atoms by various elements across the periodic table (see Table S1 and S2), where the total number of 3,119 2D HBCs are considered (Details can be found in Supplementary Information).

For the prediction of structures (members of 2D HBCs) with maximal polarization via the *ab initio* Bayesian active learning, multiple independent runs of optimization are conducted with randomly selecting initial 15 structures, and the only top five unexplored structures with the largest EI are calculated using DFT and added to each iteration (The number of added candidates can be found in Table S4). Noting that their EI values are different and their contributions to GP model are obviously different, therefore this ensures that the optimization will not fall into local optimal value. In addition, if the results of multiple independent optimization runs are almost the same, it will imply that the global optimal value is identified and the optimization process should be stopped. Contrariwise, it may be the local optimal values, and in this latter case, we need to increase the number of iterations until the results of multiple runs converge. Since each optimization run in the *ab initio* Bayesian active learning begins with randomly selecting a small number of structures as the initial training data set, we perform multiple independent optimization runs to average out the effect from the random selection. These structures are described with 11 features, which obtain *d*, MDEDM, and 9 element-related properties of constituent atoms (as listed in Table S3 and see Fig. S9(c)†), where MDEDM is the Manhattan distance of eigenvalue of the distance matrix, defined by the form of $\text{MDEDM} = \frac{1}{2}\sum_{i,j}|\lambda_i - \lambda_j|$, where $\lambda_i$ and $\lambda_j$ are the i-th and j-th eigenvalues of the distance matrix, respectively (see Table S3 for details). The polarization is related to the electron density of the materials, and the electron density depends on the atomic position, therefore, we adopted the geometric descriptor *d* and MDEDM to describe geometrical information about atomic positions of optimized geometric structures. Geometric descriptors were widely used in discovery of functional materials. [8, 13] The optimized geometric structures and electric polarizations are calculated by DFT (Supplementary

Information). As shown in Fig. 3(a), for BaPt$_2$ as an example, all 10 independent optimization runs converge to the polarization of 64.36 pC/m within 21 iterations corresponding to 120 calculated structures, indicating that the polarization value should be maximal. In addition, BaPt$_2$ does preserve good thermal and dynamical stabilities (Figs. S14 and S15†). Considering that several 2D intermetallic compounds have been confirmed experimentally, [39, 43-45] BaPt$_2$ also may be synthesized.

To check the accuracy of the method, all candidates are also calculated by DFT. It is found that maximal polarization and the corresponding structure are confirmed to be the same as those obtained by the *ab initio* Bayesian active learning, indicating a remarkable accuracy of this method. Note that after all Bayesian optimization runs are finished, all the materials are calculated to verify the accuracy of Bayesian active learning. To evaluate the computational costs of this method, we performed 10 independent optimization runs of random search for a comparison. The random search method randomly adds 5 unexplored candidates at each iteration, which can be taken as a choice of high-throughput *ab initio* calculations in actual practice.[15, 17] As illustrated in Fig. 3(b), all independent 10 optimization runs of random search come to convergence within 608 iterations corresponding to 3,055 calculated structures, which is far larger than that of the *ab initio* Bayesian active learning (120 calculated structures). Distributions of how many calculated structures are taken to find the structure with maximal polarization in 200 independent optimization runs are shown in Figs. 4(a) and (b). For the *ab initio* Bayesian active learning, the optimal structure is found within about 65 calculated structures in the most of 200 optimization runs, but for random search the distributions of calculated structures are very random. The average numbers of calculated structures of the *ab initio* Bayesian active learning and random search methods are 69 and 1618, respectively. In addition, Bayesian active learning is the feedback iterative process and it is serial in nature. High-throughput calculation could deal with a large number of materials in parallel at the same time. Nevertheless, the calculated candidates of the *ab initio* Bayesian active learning are several ten times less than that of random search, significantly reducing the computational costs with ultrahigh accuracy.

**Application in the prediction of 2D photovoltaic materials**
2D photovoltaic materials have attracted much attention due to their exciting optical and electronic properties as well as potential applications in future photovoltaic devices. Several interesting 2D photovoltaic materials have recently been reported, including 2D perovskites, [46, 47] monolayer InGeTe$_3$, [48] black arsenic–phosphorus, [49] transition-metal dichalcogenides and black phosphorus, [50] *etc*. As the monolayer NbN that belongs to the family of 2D HBCs was shown to have potential photovoltaic applications, [51] it is appealing to apply the *ab initio* Bayesian active learning to predict novel materials with promising photovoltaic properties from the family of 2D HBCs.

As the ideal band gap of single junction solar cell is around 1.5 eV, [52, 53] we may design a model of the *ab initio* Bayesian active learning to predict the 2D HBC materials whose band gaps are equal or close to 1.5 eV. In this way, we can take the expression |E$_{gap}$-1.5| as the objective function for the *ab initio* Bayesian active learning and keep this function to be minimized, where E$_{gap}$ represents the band gap. Multiple independent optimization runs are performed with random selection of 15 initial structures at the beginning, and 5 structures with the largest EI are added at each iteration (Details can be found in Table S5). When the results of multiple independent optimization runs are nearly the same, it indicates that the global optimal property is obtained and we could stop the optimization process. The only five candidates with maximum EI are calculated by DFT in each iteration. To describe these structures, we utilize the 12 features that are taken the same as the supervised machine learning model to get the band gap for a comparison (Table S3† and Fig. S9(d)†). These features include the MDEDM, *d* and 10 element-related properties of constituent atoms. The electronic band structures are calculated at Perdew-Burke-Ernzerh (PBE) [54] functional level with the on-site Coulomb interaction U [55] (Supplementary Information). All 10 optimization runs of the *ab initio* Bayesian active learning are shown in Fig. 3(c), in which the convergence of optimization runs is set to keep the function value of 0.010 eV (Counterpart is GeBi$_2$ with band gap of 1.490 eV) within the calculations of 145 structures corresponding to 26 iterations, suggesting that the band gap should be the closest to 1.5 eV.

We also calculate the electronic structures of above all HBCs by PBE+U method to show the accuracy of the *ab initio* Bayesian active learning. The results demonstrate that the band gaps close to 1.5 eV and the corresponding materials are completely consistent with the results identified by the *ab initio* Bayesian active learning. Noting that all the candidates are calculated to check the accuracy of Bayesian active learning after all optimization runs are completed. The comparison of performance between random search and the *ab initio* Bayesian active learning is shown in Figs. 3(c) and (d). It can be seen that the prediction of the structure with the ideal band gap (1.5 eV) needs to calculate up to 145 and 3,100 structures for the *ab initio* Bayesian active learning and random search, respectively, indicating that the required calculations of the former are much lower than that of the latter. Figs. 4(c) and (d) show the distributions of the numbers of calculated structures required to predict the target structure with the ideal band gap. One may observe that about 50 calculated structures are required to confirm the optimal structures in the most of 200 optimization runs via the *ab initio* Bayesian active learning, whereas for random search the distributions of calculated structures are very randomly. For the *ab initio* Bayesian active learning and random search, the average numbers of calculated structures are 70 and 1,614, respectively. Therefore, the calculated structures of the *ab initio* Bayesian active learning are only a several tenths of that of the random search, suggesting that this method could immensely decrease the computational costs and can achieve an extremely high accuracy.

To show the poor performance of the supervised machine learning in the case of unbalanced distributions of target property, we here build a supervised machine learning model for the band gap using the Gradient boosted regression (GBR)[56] algorithm for above HBCs that have unbalanced distributions of band gaps (Fig. S2(b)†). The coefficient of determination value ($R^2$) for the supervised machine learning model of band gap is very low (0.231) when all 3,119 HBC structures are selected as the training data set (Fig. S9(b)†) due to extremely unbalanced distributions of band gaps for these structures. Thus, the unbalanced distribution of target property dramatically leads to the worse accuracy of the supervised learning model, and it is very difficult to accurately find functional materials with desired property. However, the *ab initio* Bayesian active learning could be utilized to predict the materials with the band gap closer 1.5 eV from the 2D HBC family in the cases with unbalanced distributions of band gaps. Therefore, this active method provides an effective way to accelerate the prediction of advanced functional materials with desired properties, especially for unbalanced distributions of target property.

**Physical Properties of predicted 2D ferroelectric and photovoltaic materials**

Here we mainly investigate systematically all MX materials screened out by our active method by means of DFT calculations. There are 24 dynamically stable ferroelectrics with electric polarization greater than 1.0 pC/m, where their dynamical stabilities are confirmed. They are all semiconductors, including six unreported ferroelectrics (Table S8). Considering that they possess good thermodynamic and dynamical stabilities (Figs. S14 and S15 †) and 2D intermetallic semiconductors compounds $Sn_2Bi$ have been synthesized,[45] thus these materials also may be confirmed experimentally. Among these ferroelectrics, we focus on two 2D HBCs (BiAs and BiSb) with promising photovoltaic performance. We study their electronic structures and find that BiAs and BiSb possess the direct band gaps of 1.48 and 1.35 eV at the Hyed-Scuseria-Ernzerhof (HSE)[57] level, respectively (Figs. 5(a) and (b)).

A high carrier mobility is very important for the performance of photovoltaic devices. By using the deformation potential theory,[58] the electron mobility at room temperature of BiAs and BiSb are calculated to be 5335.7 and 6544.8 $cm^2V^{-1}s^{-1}$ along the zigzag direction ($x$ direction in Fig. 2), respectively, while their hole mobilities are 20.7 and 28.5 $cm^2V^{-1}s^{-1}$ along the zigzag direction ($y$ direction in Fig. 2), respectively (Table S10). Therefore, their electron mobilities are much larger than their hole mobilities, indicating an obvious electron-hole asymmetry, which is beneficial to the separation of carriers. Although the deformation potential theory does not include the longitudinal optical (LO) phonons and inter-valley scatterings that could affect the carrier mobility,[59-61] small Born effective charge and single conduction band valley at Γ point have been found in the above materials, implying weak LO phonons and inter-valley scatterings. Thus, the deformation potential theory is a reasonable method for these materials and their experimental results of mobilities may be close to the present theoretical predictions.

Furthermore, they possess the ultrahigh absorbance coefficients ($9\times10^5$ $cm^{-1}$) significantly higher than those of phosphorene, $MoS_2$, and Si by using $G_0W_0$ and BSE methods[62,63] (Figs. 5(c) and (d)). Meanwhile, their absorption covers almost the entire AM1.5G solar spectra, giving rise to high conversion efficiency in utilizing solar energy. Considering that the spin-orbit coupling (SOC) effect has an essential impact on their electronic structures whereas the optical absorption depends closely on electronic structures, we also calculate their absorption spectra by $G_0W_0$+BSE with the SOC effect. It is seen that the absorption edges are remarkably shifted toward the low-energy direction, which may be owing to the SOC effect that reduces their band gaps to 1.09 and 0.80 eV (Figs. 5(a) and (b)), respectively, and results in a wider absorption window. The reason about improvements of their absorption coefficients may be the increase of their density of states near conduction bands due to the strong SOC effect (Figs. S13 (a) and (b)†), which enables more electrons from the valence band transitioning to the conduction band. This phenomenon has also been reported in previous studies on optical properties of $TMB_2$ (TM = Cr, Mn, and Fe) materials.[64] The built-in fields existing in ferroelectrics BiAs and BiSb generate two potential differences of 1.330 and 0.691 eV between the top and bottom surfaces (Figs. S13 (c) and (d)†), respectively, which are on the same order of magnitude as that of $In_2Se_3$ (1.37 eV), [29] suggesting that the strong built-in field is favorable to decrease the carrier recombination probability. Their excellent absorbance coefficients, wide absorption range, high electron mobility and strong built-in fields make them promising candidates for ferroelectric photovoltaic devices.

The total energy in terms of the polarization $P_i$ is expressed in the Landau-Ginzburg expression [65,66] in the form of $E = \sum_i \left[\frac{A}{2}P_i^2 + \frac{B}{4}P_i^4 + \frac{C}{6}P_i^6\right] + \frac{D}{2}\sum_{<i,j>}(P_i - P_j)^2$, where $P_i$ is the polarization of the i-th unit cell, <i,j> represents the nearest neighbors, and A, B, C, D are parameters obtained by fitting to the DFT results (listed in Table S9). The first three terms are used to fit the double-well potential energy versus polarization of InP and GaP, respectively (Figs. 5(e) and (f)). To check whether the polarization could be switched by an external electric field, the energy versus polarization profiles under different vertical electric fields for InP and GaP are further investigated. The energy barriers from P↓ to P↑ polarization decrease remarkably with increasing the electric field (Figs. 5(g) and (h)). The electric field of 0.8 V/Å results in low energy barriers of 12.4 and 9.5 meV/unit cell for InP and GaP, respectively, and the critical electric fields for both should be 0.8 V/Å, which are almost the same order of magnitude as that of 2D $In_2Se_3$ (0.66 V/Å). [29, 67] Thus, the polarization of both is switchable by applying a proper vertical electric field.

## Conclusions

In conclusion, we have developed a method called as the *ab initio* Bayesian active learning that combines the active learning with the high-throughput *ab initio* calculations to highly efficiently and accurately predict functional materials with desired target property in the vast complex high-dimensional search space comprised of thousands of possible candidates. The scheme is first to construct a GPR model through a small portion of materials data from total data set, then to obtain the expected improvement acquisition function of unexplored data set based on the predictions of GPR model and to select the unexplored data with the maximized expected improvement acquisition function and add them into the training dataset for retraining new GPR model, and finally to perform multiple iteration optimizations to get a globally optimized value. It is noted that the maximum of the acquisition function corresponds to the best average predicted property values ($\mu(x)$) and maxima of uncertainty ($\sigma(x)$). This method was applied to the 3,119 2D HBCs, and the results demonstrate that a number of materials with maximal polarization and proper photovoltaic band gaps are accurately identified based on the calculations of only a few tenths of the number of candidates in comparison to random search, resulting in a significant reduction of computational costs. Several unreported stable ferroelectric and photovoltaic semiconductors with good performance are thus obtained. A crucial character of this method is that it is independent of the specific materials, target properties and the distributions of materials data, and is very suitable for predicting the desired functional materials in the cases with unbalanced distributions of target property. It is expected that this efficient and accurate approach could generally be applied to predict more other classes of advanced functional materials with desired target properties at ultralow computational costs and ultrahigh accuracy.

## Computational methods

### First-principles calculations

The first-principles calculations are performed based on the density functional theory (DFT) implemented in the Vienna *Ab initio* Simulation Package (VASP) [68] within projected augmented wave (PAW) method.[69] Perdew-Burke-Ernzerhof generalized gradient approximation (PBE-GGA) [54] for exchange-correlation potential was adopted. A cutoff of 450 eV was chosen, and the Brillouin zone was sampled using Γ-centered

Monkhorst–Pack k-point grids of 12×12×1. Optimized atomic structures were achieved when the force on all atoms and the energy were converged to 0.01 eV/Å and $10^{-6}$ eV, respectively. All electronic structures are calculated using the PBE-GGA + U method. We apply Hubbard U corrections to account for Coulomb correlated potential in 3d, 4d and 5d transition metals, which were chosen as U = 4 eV, 2.5 eV and 0.5 eV, respectively, which are usually reasonable for them.[70, 71] See supplemental information for more details.

**Gaussian process regression**

The Gaussian process regression (GPR)[21] is adopted as the surrogate model in active learning, which is in the open-source scikit-learn package.[72] Gaussian process (GP) is a collection of random variables satisfying a joint Gaussian distribution. The targeted functions f(x) can be described by using the Gaussian process, which has the form $f(x) \sim GP(m(x), k(x, x'))$ where x and $x'$ are input feature vectors of two different data, m(x) and $k(x, x')$ are their mean function and kernel function, respectively. GPR is a non-parametric regression technique, which builds a distribution of functions that are accord with the training data set. An n-dimensional squared exponential kernel is adopted as the kernel function, which has the form $k(x, x') = \exp\left\{-\frac{\sum_{i=1}^{n} |x_i - x_i'|^2}{\sigma_i^2}\right\}$, where x and $x'$ are input features vectors of two different data, $x_i$ is the i-th feature of n-dimensional input feature vectors for each data, n is the dimensional of input feature vectors of each data, and $\sigma_i^2$ is a hyper-parameter obtained using the maximum likelihood estimate.

## Author Contributions

## Conflicts of interest

The authors declare no competing interests.

## Acknowledgments

This work is supported in part by the National Key R&D Program of China (Grant No. 2018YFA0305800), the Strategic Priority Research Program of CAS (Grant No. XDB28000000), the NSFC (Grant No. 11834014), Beijing Municipal Science and Technology Commission (Grant No. Z191100007219013), and the Fundamental Research Funds for the Central Universities. The calculations were performed on Era at the Supercomputing Center of Chinese Academy of Sciences and Tianhe-2 at National Supercomputing Center in Guangzhou.

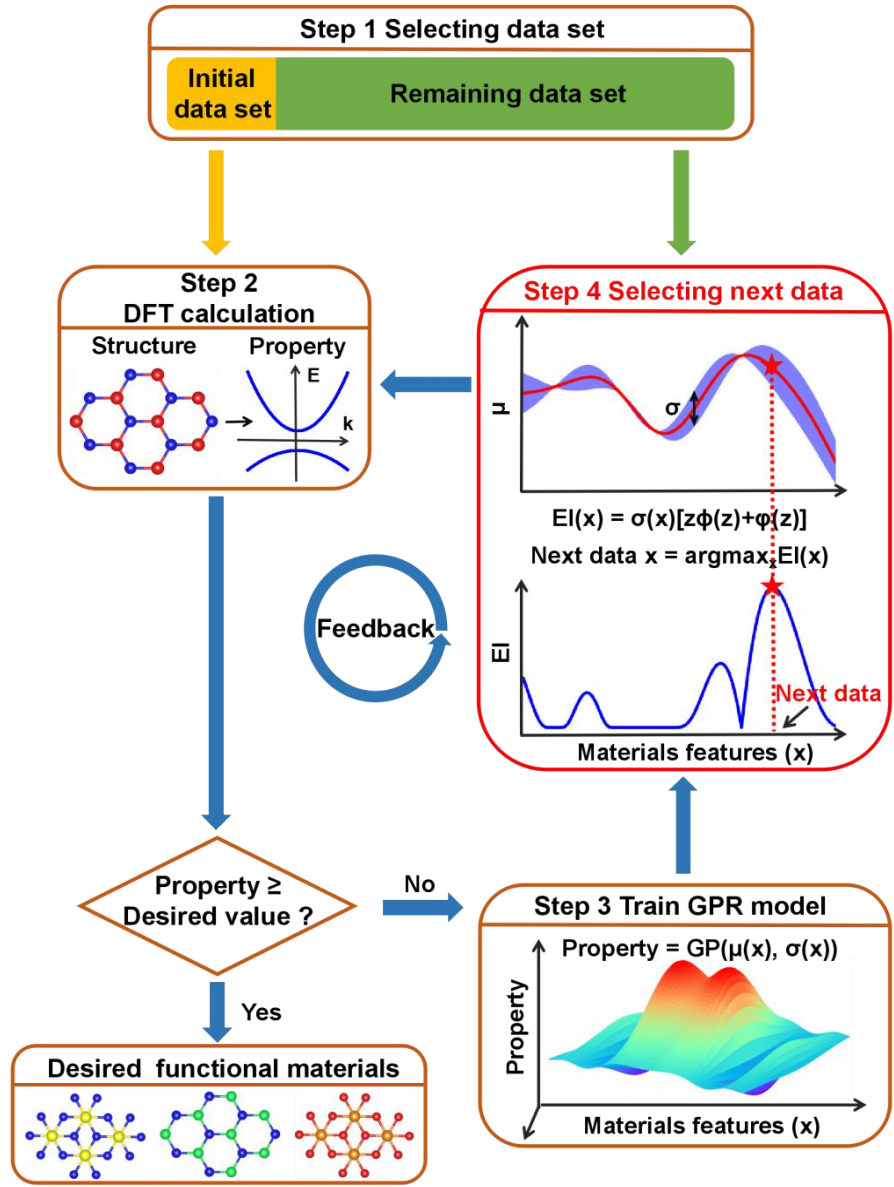

Fig. 1 The schematic flowchart of the *ab initio* Bayesian active learning for the prediction of target functional materials. The yellow and green cylinders represent initial data set and remaining data set, respectively. The next new data are iteratively selected and added into the initial data set by maximizing the expected improvement (EI) based on the Gaussian process regression (GPR) model-predicted mean target properties (μ) and standard deviation values (σ) obtained by the GPR model. The target property can be described by using the Gaussian process (GP), which is a collection of random variables satisfying a joint Gaussian distribution, x is input feature vectors of materials data, μ(x) and σ(x) are their mean predicted properties and standard deviation values (uncertainties) of the joint Gaussian distribution,

respectively. $\Phi(z)$ and $\varphi(z)$ are cumulative density function and probability density function of standard normal distributions, respectively. $z = (\mu(x) - f^{max})/\sigma(x)$ and $z = (f^{min} - \mu(x))/\sigma(x)$, where $f^{max}$ and $f^{min}$ are the maximum and minimum actual values in training data set, respectively. The height of the blue shaded area represents the uncertainties ($\sigma$). The red star represents the selected next data.

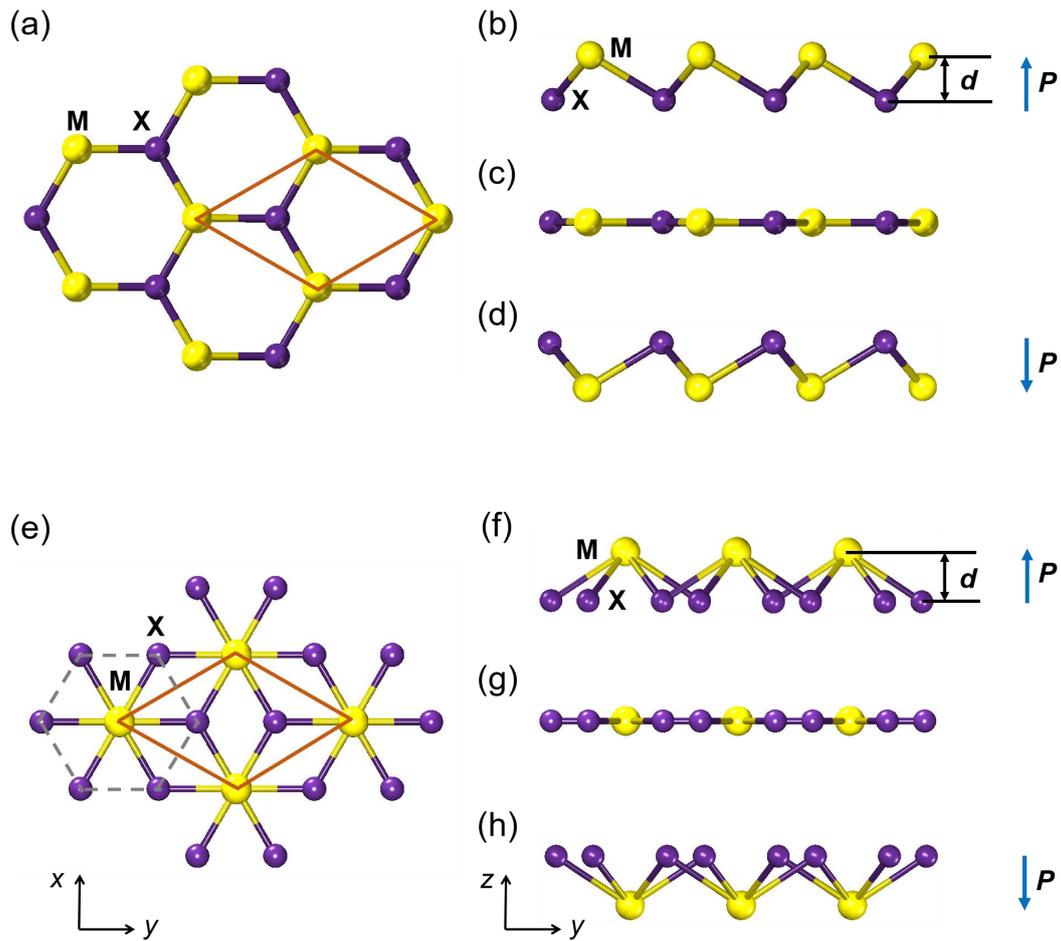

Fig. 2 The schematic structures of 2D hexagonal binary compounds (MX and MX$_2$, M and X atoms are different atoms). (a)(e) Top view. The orange parallelogram indicates the unit cell. Gray dash lines comprise of a honeycomb lattice of MX$_2$ formed by X atoms. (c)(g) Side view of paraelectric phases of (c) MX and (g) MX$_2$. (b)(d)(f)(h) Side view of ferroelectric phases of (b)(d) MX and (f)(h) MX$_2$ with different polarization directions. $d$ represents the vertical relative displacement between M and X atoms. The blue arrows (up/down) indicate the directions of out-of-plane electric polarization ($P$). Note that the directions of $P$ of MX structures may not represent the actual directions due to two equivalent atoms M and X. The yellow and purple balls represent M and X atoms, respectively.

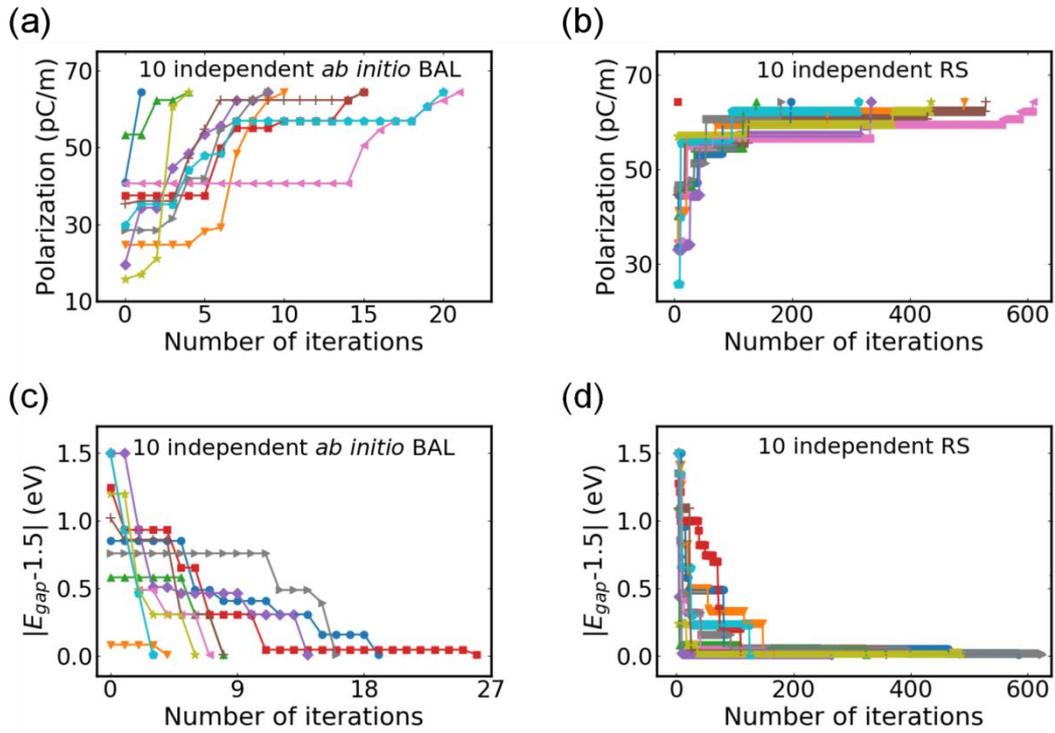

Fig. 3 Comparison between the *ab initio* Bayesian active learning (BAL) and random search (RS) for 10 independent optimization runs. The 10 independent optimization runs are conducted with randomly selecting initial 15 structures, and the top five unexplored structures with the maximal expected improvement are added to each iteration for predicting the optimal structure with (a)(b) maximal electric polarization and (c)(d) band gap close to 1.5 eV, respectively.

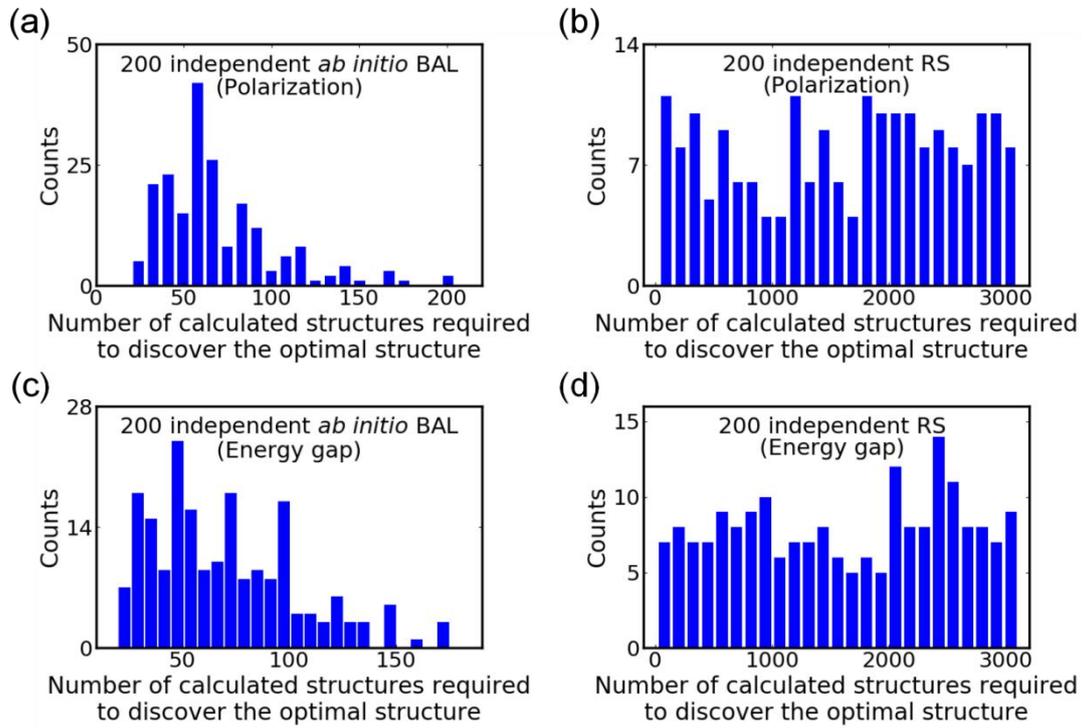

Fig. 4 Comparison between the *ab initio* Bayesian active learning (BAL) and random search (RS) for 200 independent optimization runs. Histogram of the number of calculated structures required to predict the optimal structure with (a)(b) maximal electric polarization and (c)(d) band gap close to 1.5 eV.

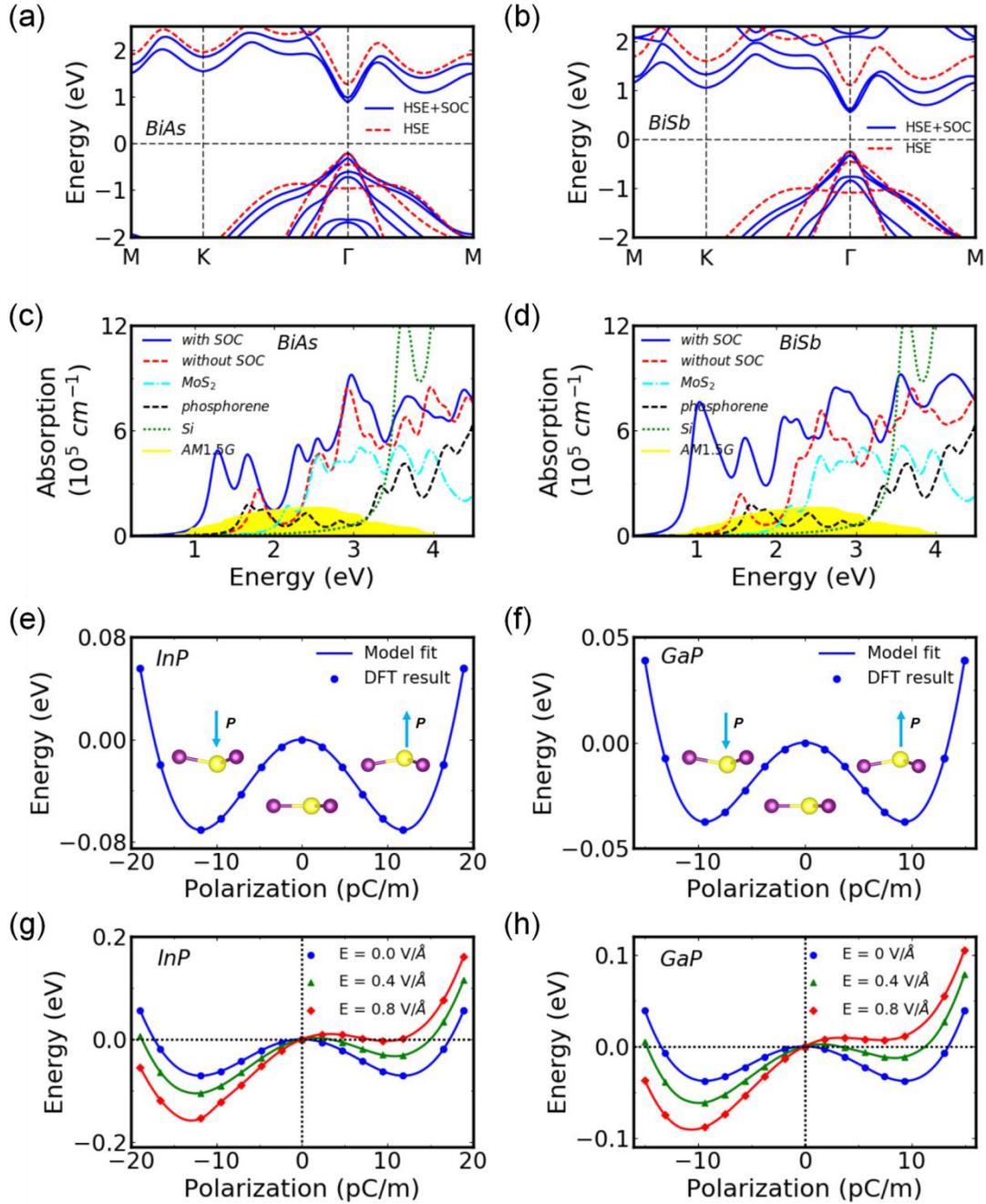

Fig. 5 (a)(b) Band structures at the HSE (red dash line) and HSE+SOC level (blue line) of two photovoltaic materials (BiAs and BiSb). (c)(d) Absorption spectra using $G_0W_0$+BSE method with/without SOC for BiAs and BiSb. The yellow background represents the incident AM1.5G solar flux. (e)(f) Double-well potential energy versus polarization of two ferroelectrics (InP and GaP). Blue points and blue curves are the DFT-calculated total energies and fitted curves with Landau-Ginzburg model, respectively. Cyan arrows represent the directions of polarization (***P***) in ferroelectric phases. (g)(h) The energy versus polarization of two ferroelectrics InP and GaP by applying different vertical electric fields. Points and curves are the DFT-calculated total energies and fitted curves, respectively.